\documentstyle[12pt]{article}

\begin{document}
\topmargin=-.5in
\oddsidemargin=.1in
\evensidemargin=.1in
\vsize=23.5cm
\hsize=16cm
\textheight=23.0cm
\textwidth=16cm
\baselineskip=24pt

\hfill{ITP-SB-92-08}\\
\smallskip
\hfill{May 1992}

\vspace{1.0in}

\centerline{\Large \bf SOLITONS WITH INTEGER FERMION NUMBER}

\vspace{.85in}

{\baselineskip=16pt
\centerline{\large Alfred S. Goldhaber}
\bigskip
\centerline{\it Institute for Theoretical Physics}
\centerline{\it State University of New York}
\centerline{\it Stony Brook, NY 11794-3840}}

\vspace{.75in}

\centerline{\Large Abstract}

\vspace{1.0in}

Necessary and sufficient conditions are found for any object in $3+1$
dimensions to have integer rather than fractional fermion
number.
Nontrivial examples include the Jackiw-Rebbi monopole and the already well
studied Su-Schrieffer-Heeger soliton, both displaying integer multiples of
elementary charges in combinations that normally are forbidden.

\newpage

        Jackiw and Rebbi in `Solitons with fermion number $1/2$'$^1$
introduced
the concept of fractional charge.  The aim of the following analysis is to
define the limits of the fractional
charge domain for objects in $3 + 1$ dimensions.

Fractional charge only is
remarkable if it is an eigenvalue of the
corresponding quantum observable, rather than just an expectation value --
it must be sharp to be significant.  The conditions for a
localized charge to be sharp have been investigated for systems in $1 + 1$
dimensions,$^2$ where the definition requires a spatially smoothed
weighting
of the corresponding charge density operator.  In higher dimensions
temporal
smoothing or a frequency cutoff also is required.$^3$  This means that
without an energy threshold for charge-carrying excitations it may,
(sometimes
even must$^4$) be
impossible to define a sharp charge, since the frequency cutoff does not
preclude long-wavelength fluctuations in the charge density.

JR$^1$ considered two examples of a
quantum Fermi field coupled to a specific classical, static Bose field
configuration (a  soliton).  In both cases the Dirac equation
for the
Fermi field
includes a single mode with zero frequency, and the rest of the spectrum is
completely symmetrical between positive and negative frequencies.
Consequently, the state in which the zero frequency mode is occupied and
that in which it is not ought to be charge conjugate to each other.  Since
they differ by one unit in fermion number $F$, they should be characterized
by
$F = \; \pm 1/2$.  For the first example, where the Fermi
field is Yukawa coupled to a sine-Gordon soliton in $1+1$ dimensions, the
treatment of Fermi
and Bose fields was made symmetrical by bosonization of the Fermi
field.
This
allows back-reaction of the fermion on the boson degrees of freedom,
and confirms the suggestion that the object carries half-integer $F$.$^5$

Su, Schrieffer, and Heeger$^6$ independently discovered a system
with fermion zero modes leading to JR oscillators, except that in
their model (of polyacetylene) there are two species of fermion, electrons
with spin up and electrons with spin down.  This fermion doubling replaces
fractional charge with charge \underline{dissociation} -- an SSH soliton
may
exist in
any of four nearly degenerate states, characterized by $F = \; \pm 1$, spin
$S = 0$
(spin singlet), or $F = 0, S = 1/2$ (spin doublet).  The two charges
(electron
number and spin) which characterize a free electron have been `torn apart'
so
that soliton states carry one or the other but not both.

     The concept of fractional charge has blossomed since JR.  A
soliton which lacks charge conjugation symmetry may possess fractional
charges other than $1/2$, whether rational$^7$ or irrational.$^{8,9}$
Perhaps the
most striking experimentally studied example came from defining$^{10}$ a
charge somewhat different from
the
localized charges mentioned above, the `capacitor charge' of a weak link in
a conducting circuit, $\tilde{Q} = CV$.  Here $V$ is the
electric
potential across the link, and $C$ is the
classical electric capacitance of the link.  If
$\tilde{Q}(t)$ is
sharp
then electrons should jump across the link at sharply defined
times separated by the unit $e/I$, with $e$ the electron charge and $I$
the mean current, a prediction$^{10}$ verified by experiment.$^{11}$
Provided a
suitable frequency cutoff is introduced, $\tilde{Q}$ indeed should be
sharp,
even if the conductor is a normal metal which has zero energy threshold for
charge-carrying excitations.  This is possible because charge density
fluctuations which
produce a significant shift in $\tilde{Q}$ must be localized near the link,
and
do have a nontrivial energy threshold.

         Since at least the potentiality for fractional charge is by now
seen as ubiquitous, it becomes interesting to ask when is it
\underline{not} found, i.e., what are the
conditions
for a soliton to carry charges or quantum numbers with integer rather than
fractional values?  Let us begin to attack this question by recalling the

\underline{Theorem of Jackiw and Schrieffer}:$^{12}$  A soliton whose
spectrum is
invariant
under a charge conjugation symmetry $C$ which reverses the sign of $F$ may
have
integer
$F$, or half-integer $F$, but no other fractional value is allowed.

\underline{Proof}:  For an isolated soliton, the only way $F$ (assumed
conserved) can
change is by scattering processes in which the number of fermions incident
differs
from the number emerging.  This means that two allowed values of $F$ must
differ by an integer.  $C$ symmetry implies that for every
allowed $F$, $-F$ also is allowed.  Therefore the difference $2F$ must be
an
integer, and $F$ must be either an integer or a half-integer.  If one
allowed
 $F$ is an integer (half-integer), so must be all the others, since
they differ by integers.  QED

        This theorem shows that to answer our question we need find
only what conditions must supplement charge conjugation symmetry in order
to
exclude half-integer eigenvalues. It also suggests a strategy for
determining those conditions.  If absorbing a fermion changes other
conserved charges of the soliton as well as $F$, then extra consistency
requirements may follow.  Therefore let us proceed by attending to other
charges carried by fermions.  The problem naturally divides solitons into
two classes, (A) magnetic monopoles (interacting with fermions whose
electric charge has the minimum magnitude $e$ allowed by the Dirac
quantization
condition$^{13}$), and (B) all others.

        The reason for this division is that according to the
spin-statistics connection a fermion in $3 +1$ dimensions must carry
half-integer spin.  However, for a minimal electric charge in the field of
a magnetic monopole there is an extra electromagnetic angular momentum
which also is half-integer.  Consequently, it is possible for the
monopole to absorb a fermion with the appropriate magnitude of electric
charge without absorbing angular momentum.  For any
other kind of soliton, or for a monopole interacting with fermions whose
electric charges are even multiples of the minimum Dirac unit, absorbing a
fermion requires the absorption of a half-integer unit of angular momentum.
 Therefore spin is a suitable candidate for the extra quantum number
associated with depositing a fermion on a soliton for all cases in category
B.  For category A, the only apparent (and inevitably open) option for the
extra quantum number
is
electric charge.  Let us now analyze cases A and B separately.

\underline{Integer $F$ Theorem A}:  If the electromagnetic vacuum angle
$\theta$ vanishes, then a magnetic monopole symmetric under fermion
conjugation $C$ must carry integer $F$.

\underline{Proof}: Let us begin by making explicit several assumptions.
Take $C$ to reverse electric charge $Q$ as well as $F$, since the proof
becomes trivial otherwise.  Restrict attention only to fermions with $Q$
(measured in units of the elementary fermion electric charge $e$) an odd
integer, since the alternative will be
covered under Theorem B, to follow.  Treat both $Q$ and $F$ as conserved,
localizable, sharp quantum observables.
By the method of the Jackiw-Schrieffer theorem, the only possible values of
$Q$ and $F$ are integers or half-integers, and furthermore both must be in
the same class.  This follows because adding one fermion
changes both $Q$ and $F$ by an odd integer.  Thus, if $F$ is a
half-integer, so is $Q$.  The proof is reduced to showing that $Q$ must be
an integer.

To establish this, let us recall the basis for the Dirac
quantization condition.  Dirac's discovery of the condition exploited the
gauge invariance of electromagnetism,$^{13}$ and could be phrased by
saying
that the monopole is described formally as one end of an infinitely
thin magnet.  In order that this magnetic line, or Dirac string, be
unobservable, the
Aharonov-Bohm phase$^{14}$ for diffraction of electrically charged
particles
on either side of the line should be an integer multiple of $2\pi$, and
this implies the quantization condition
\begin{equation}
qg = N \hbar c/2 \; ,
\end{equation}
where $q$ is the electric charge, $g$ is the magnetic pole strength, and
$N$
is an integer.

A second way to obtain the same condition is by insisting on
proper quantization of the total angular momentum of a charge-pole system.
At the simplest level, the argument is that the Thomson electromagnetic
angular momentum$^{15} \; qg/c$ along the line from charge to pole should
be
quantized.$^{16}$
This suggests a possible escape from Dirac's condition for the case of
dyons carrying both electric and magnetic charge:  The generalized
condition$^{17}$
\begin{equation}
q_1g_2-q_2g_1 = N \hbar c/2 \; ,
\end{equation}
where the subscripts refer to the charges on the respective particles, has
as a possible solution that each dyon carries a fractional electric
charge proportional with a universal ratio to its magnetic charge.  This
makes the Thomson angular momentum of a dyon pair vanish identically, so
that its quantization gives no further constraint.

However, if the
introduction of these fractional charges is represented by a term in the
action density proportional to $K_{\mu}A_{\mu}$, where $K$ is the monopole
current and $A$ is the usual electromagnetic four-potential, then the $q_i$
still must obey Dirac's original condition.  The reason is seen most
easily in terms of Dirac strings.  The fact that there is no
velocity-dependent `magnetic' force between two such dyons means that in
any simply connected region of the relative coordinate the vector potential
may be written as a pure gradient. Still we
need to check for an observable Aharonov-Bohm phase when the relative
motion
corresponds to diffraction around what is now an endless Dirac string.  To
avoid a nontrivial phase, Dirac's condition Eq. 1 is required.$^{18}$

The fact that the straightforward or obvious formulation of possible
fractional dyon charges does not work should be no surprise, since one is
giving mathematical expression to the placement of charges `by hand' on
each monopole,  It seems natural that Dirac's condition forbidding
fractional
charges at arbitrary locations should apply also to the limiting case when
their locations are made to coincide with those of monopoles.
Nevertheless,
there is  a unique, consistent way to introduce fractional charge, and that
is by modifying electrodynamics to include a nontrivial vacuum angle
$\theta$.$^{8,19}$  Such an angle is believed to be a possible result of
instanton tunneling effects, and is represented in the action density by a
term proportional to $\theta{\bf E \cdot B}.^{20}$

That this leads to fractional dyon charge can be seen directly from the
equations of motion.$^{21}$  That it avoids the breaking of gauge
invariance
inevitable with `hand placement' has been shown to be due to the appearance
of a
second gauge-noninvariant term which exactly cancels the one coming from
the $K_{\mu}A_{\mu}$ coupling.$^{19}$  This result is easily understood,
since the $\theta$ term in the action is manifestly gauge invariant without
the necessity of an integration by parts, and therefore must produce only
gauge invariant results.  Provided only that there exists a gauge
invariant, local, low-energy effective action describing charges and poles
interacting with the electromagnetic field, the $\theta$ term is the unique
mechanism for introducing fractional dyon charge, since it must include
$K_{\mu}A_{\mu}$ and be gauge invariant.  Any proposed alternate
could be rewritten as the $\theta$ term plus another gauge invariant piece
not including $K_{\mu}A_{\mu}$ and therefore irrelevant to fractional
charge.

We see that for vanishing $\theta$ there can be no fractional $Q$
on a monopole, but that means there can be no fractional $F$, and Theorem A
is proven.  Before going on to the much more straightforward Theorem B, let
us pause to examine some related issues.

1) \underline{Apparent confirmations of $F=1/2$ for the JR monopole}.
Following the suggestion of JR that the zero mode which they discovered for
isospinor fermions coupled to the gauge and Higgs fields of an 't
Hooft-Polyakov monopole$^{22}$ implies $F=1/2$, the same $F$ value was
obtained by explicitly field-theoretic methods.$^9$  However, the formalism
behind this result was based on quantization of the Fermi fields only, with
the Bose fields treated as adjustable classical backgrounds.  Thus it
amounts to a confirmation of the original JR calculation, but does not
account for possible back-reaction of the fermion on the boson degrees of
freedom.

The issue of back-reaction was attacked in an approach which bosonized the
Fermi
fields, introducing plausible boundary conditions for the bosonized fields
at the core of the monopole.$^{23}$  This charge-conjugation-symmetric
approach inevitably produced $Q = 1/2$ along with $F = 1/2$, and therefore
included an implicit assumption of nontrivial vacuum angle.  If we reject
that assumption we must revise the boundary conditions, no matter how
plausible they may appear, and are left with no evidence for fractional
$F$.

2) \underline{Counting states}.  If JR's disarmingly simple argument does
not yield the required integer $F$, where is the the additional
`duplexity'$^{24}$ which compensates or conceals the $\pm 1/2$ which they
found?  To give this the direct answer it deserves, let us repeat the JR
analysis$^{1,12}$ in algebraic
form.  Commuting with the monopole Hamiltonian are three
operators,
$\psi_0$ -- the
Hermitean part of the operator in the expansion of the Dirac field
corresponding to the
JR fermion zero mode, $C\!P$ -- the relevant charge conjugation symmetry,
and $F$.  $F$ anticommutes with $C\!P$, and is raised or lowered one unit
by
$\psi_0$, which in turn commutes with $C\!P$. The minimum nontrivial
representation of this algebra is two dimensional, and the basis states may
be chosen as $|F = \pm 1/2>$.

Now we are ready for the missing ingredient:  In the classical limit
electric charge $Q$ need not be sharp, but the charge signature operator $U
= (-1)^Q$ still has definite, conserved eigenvalues. In the absence of
$C\!P$-violating effects these eigenvalues are $U = e^{2\pi iT_3}$.$^8$
   Here $T_3$ is the generator of rotations about the abelian or
massless-photon direction in the $SO(3)$ gauge space, implying $e^{2 \pi
iT_3}= \pm 1$.  $U$ evidently commutes with $C\!P$ and $F$, but
anticommutes
with
$\psi_0$.  If the sector with $U = +$ contains a state with nonzero $F$,
then
the $C\!P$ conjugate state with $-F$ must differ by an even integer.  Thus
the
minimum nontrivial $U = +$ multiplet must contain two states with $F =\pm
1$.
Since this yields a two dimensional representation of $C\!P$, there must
also
be a two dimensional $F = 0$ representation for $U = -$, yielding a
collection
of four states as the smallest nontrivial multiplet.  Although $Q$ is not
sharp in this basis, the two
states for $U = -$ may be chosen to carry the expectation values $<Q> \; =
\pm 1$.

Of course it still is possible to choose two $C\!P$ conjugate states which
differ in $F$ by one unit, and therefore have $<F> \; =\pm 1/2$.  In other
words, by ignoring $U$ we may repeat the original JR analysis with no error
except that the charge conjugate states no longer are unique, and hence the
fractional expectation values are not
eigenvalues.

Upon going to a basis with $Q$ sharp, which is allowed in the classical
limit, and necessary if quantum corrections are to be included, we obtain
as
the natural structure for the multiplet of smallest mean-squared charge a
quartet, one doublet with $F = \pm 1, Q = 0$, and one doublet with $F = 0,
Q
= \pm 1$.  All this is in perfect analogy with the SSH soliton,$^6$ so that
the JR monopole actually is an example of charge dissociation rather than
fractional charge.

3) \underline{Necessity of the $\theta = 0$ hypothesis}?  The assumption of
vanishing vacuum angle is sufficient to prove Theorem A.  It also is
necessary, if the example of Dirac electrons interacting with a point Dirac
monopole is considered.  For that case a suitable $\theta$ can be
found$^{25}$ to make the monopole an ideal JR oscillator, whose degenerate
ground states carry $F = Q = \pm 1/2$.  For the nonsingular, and therefore
perhaps
more natural, 't Hooft-Polyakov monopole, our question splits into two:
Could the fermion-Higgs coupling somehow induce a nontrivial $\theta$?
Even if there is such an angle, will the monopole acquire fractional $F$?

The first question describes a logical possibility, but I am unaware of any
mechanism other than instanton tunneling which might produce nonzero
$\theta$, so that the burden of proof should lie on proponents of such a
notion.  The answer to the second question seems more straightforward.
Since $\theta$ is linked to $Q$, cranking $\theta$ from 0 to $\pi$
should take $Q$ through fractional values from 1 to 0, but should not
influence $F$.  Indeed, one can follow the spectrum of single-particle
bound states associated with such a sequence. At $\theta = 0$ there are
effectively two zero modes, one with positive and one with negative
electric charge, yielding four nearly degenerate states with  $F = \pm 1,
Q = 0;
F = 0, Q =\pm 1.$   These are the states I have argued one should expect
from the JR mechanism.  Between $\theta = 0$ and $\theta =\pi/2$ the
positive charge bound state rises from $E = 0$ to $E = mc^2$ (and then
disappears into the continuum), while the negative charge state descends
from $E = 0$ to $E = -mc^2$.  At $\theta =\pi$ the monopole has a single
ground state with $F = Q = 0$. Thus it appears that adjusting $\theta$
could undo the twice-doubled degeneracy associated with the JR mechanism in
this setting, and could produce fractional $Q$, but it could not produce
fractional $F$.

For all other solitons, including all condensed-matter
defects, we have

\underline{Integer $F$ Theorem B}:  An object self-conjugate under a
unitary charge conjugation symmetry $C$, with fermion number and spin both
sharp, must carry integer $F$.

\underline{Proof}:  Again there are assumptions which should be made
explicit.  We are omitting the case of Theorem A, so that adding a fermion
to the object does change its spin by a half-integer.  $C$ is assumed to
commute with rotations, so that the unitarity  of $C$ implies its
commutation with angular momentum and spin.  Therefore charge conjugate
states must have the same spin.  This means that $2F$ must be an even
integer, since otherwise the states would differ in spin by a half-integer.
 Hence $F$ is an integer -- the proof is complete.  Let us now look for
counterexamples if assumptions are violated.

1) \underline{Antiunitarity}.  If $C$ is antiunitary then it anticommutes
with spin, and the above proof breaks down.  A possible example is the SSH
model with a very strong magnetic field along the polymer chain direction,
so that the valence parallel-spin electron band is completely
nonoverlapping with the antiparallel band.  Then a soliton in a
`quarter-filled' background (i.e., the parallel band half-filled, the
antiparallel unfilled) should have $S_z = F/2 = \pm 1/4$.  Since rotational
symmetry has been broken from $O(3)$ to $O(2)$, there is no intrinsic
contradiction in a fractional eigenvalue for $S_z$.  This is different from
the monopole case discussed earlier -- where $Q = 2T_3$ must be an integer
-- because there the isospin gauge symmetry (even though hidden by the
Higgs
mechanism) remains unbroken.

2) \underline{Spin delocalization}.  If spin is not localizable because of
strong and long-range magnetic correlations, then adding a fermion need not
change even the expectation value of spin -- no spin is deposited locally.
Consequently the difference in $F$ between charge conjugate states can be
an odd integer, without any difference in the expectation value of spin.
Precisely this may occur in certain organic charge-transfer salts, yielding
$F =\pm 1/2, <S_z> \; = 0$, contrary to the previous claim of fractional
spin
for such a system.$^{26}$

We have seen that zero vacuum angle, localizability of $F$, $Q$ and $\bf
S$, and obedience to a unitary $C$ symmetry are the general necessary and
sufficient conditions that a soliton in $3 + 1$ dimensions carry integer
$F$.
The premier application and inspiration of this analysis is the subtle
Jackiw-Rebbi monopole,$^1$
which helped reveal a significant new phenomenon by evoking fractional
charge even without possessing it.

   I have benefited from discussions with many people, especially Hidenaga
Yamagishi, who early and consistently focused on the crucial role of the
vacuum angle, and Steven Kivelson, who shared an education on the
significance of sharp charge.  This work has been supported in part by the
National Science Foundation, Grant PHY \#90-8936.

\newpage

\centerline{\Large References}

\begin{enumerate}
 \item  R. Jackiw and C. Rebbi {\it Phys Rev D} {\bf 13} 3398 (1976).
 \item  S. Kivelson and J.R. Schrieffer {\it Phys Rev B} {\bf 25} 6447
(1982), R. Raraman and J.S. Bell {\it Phys Lett} {\bf 116B} 151 (1982); and
ref. 4 in second citation of ref. 3 below.
 \item  M. Requardt {\it Commun Math Phys} {\bf 50} 259 (1976); A.S.
Goldhaber
and S.A. Kivelson {\it Phys Lett} {\bf 255B} 445 (1991).
  \item  A.S. Goldhaber in \underline{Workshop on Foundations of Quantum
Mechanics} eds. T.D. Black, M.M. Nieto, H.S. Pilloff, M.O. Scully and R.E.
Sinclair (World Scientific, Singapore 1992).
  \item  R. Shankar and E. Witten {\it Nucl Phys} {\bf B141} 349 (1978);
{\bf
B148} 538 (E) (1979); E. Witten {\it ibid} {\bf B142} 285 (1978).
  \item  W.P. Su, J.R. Schrieffer and A.J. Heeger {\it Phys Rev Lett} {\bf
42}
1698 (1979).
  \item  W.P. Su and J.R. Schrieffer {\it Phys Rev Lett} {\bf 46} 738
(1981).
  \item  E. Witten {\it Phys Lett} {\bf 86B} 282 (1979).
  \item  J. Goldstone and F. Wilczek {\it Phys Rev Lett} {\bf 47} 986
(1981).
  \item  D.V. Averin and K.K. Likharev {\it J Low Temp Phys} {\bf 62} 345
(1986).
  \item  P. Delsing, K.K. Likharev, L.S. Kuzmin and T. Claeson {\it Phys
Rev
Lett} {\bf 63} 1861 (1989).
  \item  R. Jackkiw and J.R. Schrieffer {\it Nucl Phys} {\bf B190} 253
(1981).
  \item  P.A.M. Dirac {\it Proc R Soc London} {\bf A133} 60 (1931).
  \item  Y. Aharonov and D. Bohm {\it Phys Rev} {\bf 115} 485 (1959).
  \item  J.J. Thomson {\it Phil Mag} {\bf 8} 331 (1904).
  \item  M.N. Saha {\it Phys Rev} {\bf 75} 1968 (1949); H.A. Wilson
{\it ibid} 309.
  \item  J. Schwinger {\it Phys Rev} {\bf 173} 1536 (1968); D. Zwanziger
{\it
Phys Rev} {\bf 176} 1480 (1968).
  \item  A.S. Goldhaber {\it Phys Rev Lett} {\bf 36} 1122 (1976).
  \item  F. Wilczek {\it Phys Rev Lett} {\bf 48} 1146 (1982); A.S.
Goldhaber,
R. MacKenzie and F. Wilczek {\it Mod Phys Lett A} {\bf 4} 21 (1989).
  \item  C.W. Bernard and E.J. Weinberg {\it Phys Rev D} {\bf 15} 3656
(1977).
  \item  S. Coleman in \underline{The Unity of Fundamental Interactions}
ed. A.
Zichichi (Plenum, New York 1983) p 21.
  \item  G. 't Hooft {\it Nucl Phys} {\bf B79} 276 (1974); A.M. Polyakov
{\it
JETP Lett} {\bf 20} 194 (1974).
  \item  C.G. Callan Jr. {\it Phys Rev D} {\bf 26} 2058 (1982).
  \item  A.S. Goldhaber {\it Phys Rev D} {\bf 33} 3697 (1986).
  \item  H. Yamagishi {\it Phys Rev D} {\bf 27} 2383 (1983); B. Grossman
{\it
Phys Rev Lett} {\bf 50} 464 (1983).
  \item S.C. Zhang, S. Kivelson and A.S. Goldhaber {\it Phys Rev Lett} {\bf
58} 2134 (1987).
\end{enumerate}
\end{document}